\documentclass[twocolumn,preprintnumbers,showpacs,showkeys,amsmath,amssymb,epsf]{revtex4}
\usepackage{graphicx}
\usepackage{dcolumn}
\usepackage{bm}
\begin{document}
\title{Bose-Einstein condensation and a two-dimensional walk model}
\author{Farhad H. Jafarpour}
\email{farhad@ipm.ir}
\affiliation{Physics Department, Bu-Ali Sina University, 65174-4161 Hamedan, Iran}
\date{\today}
\begin{abstract}
We introduce a two-dimensional walk model in which a random walker can only move on the first quarter of a two-dimensional plane. We calculate the partition function of this walk model using a transfer matrix method and show that the model undergoes a phase-transition. Surprisingly the partition function of this two-dimensional walk model is exactly equal to that of a driven-diffusive system defined on a discrete lattice with periodic boundary conditions in which a phase transition occurs from a high-density to a low-density phase. The driven-diffusive system can be mapped to a zero-range process where the particles can accumulate in a single lattice site in the low-density phase. This is very reminiscent of real-space Bose-Einstein condensation.
\end{abstract}
\pacs{05.70Fh, 05.70.Ln,05.50.+q}
\keywords{Walk model, driven-diffusive system, non-equilibrium phase transition}
\maketitle
\section{Introduction}
Over the last couple of years there has been a growing interest in studying of the connections between the one-dimensional driven-diffusive systems and the two-dimensional walk models \cite{BGR04,BE04,BCEP06,BJJK09}. It has been shown that a properly defined steady-state normalization factor, called the partition function, of some of the one-dimensional driven-diffusive systems with open boundaries obtained using a matrix product method (reader can see \cite{BE07} for a review) is equal to the partition function of a two-dimensional walk model obtained using a transfer matrix method \cite{BGR04,JZ01,JZ02}.\\
It is known that the one-dimensional driven-diffusive systems exhibit a variety of interesting critical behaviors, such as non-equilibrium phase transition and real-space condensation, by changing their microscopic reaction rates. In \cite{BE03} the authors have shown that the phase transitions in the steady-state of non-equilibrium systems can be investigated and classified using the Lee-Yang theory of phase transitions. The asymmetric simple exclusion process (ASEP) is known as the simplest system among the one-dimensional driven-diffusive systems \cite{DEHP}. In an inhomogeneous version of this process (also known as the ASEP in the presence of the impurities or second class particles) the particles hop in a preferred direction with different hopping rates on a one-dimensional lattice with periodic boundary conditions \cite{DEHP,M96,E96,E00}. It has been shown that there is an exact mapping from this process to a zero-range process in which the particles behave like bosons and the steady-state is a product measure \cite{E00}. The zero-range process was first introduced into the mathematical literature as an example of interacting Markov processes. By fine-tuning of the microscopic hopping rates, a finite fraction of a system's mass can accumulate within a microscopic region of the system. This can be thought of as a traffic jam. In an equivalent zero-range process, it turns out that the average number of the particles at a single lattice site in this phase can be of order of the system size. This is in contrast to the exclusion process where each lattice site can be only occupied by a single particle. The condensation in this phase reminds us of real-space Bose-Einstein condensation \cite{E96,E00}.\\
In recent years, different types of walk models have been introduced where some of these models can describe physical phenomena such as polymer phase transitions. Some examples are given in \cite{BOR08}. As we mentioned earlier, two-dimensional versions of these models have been adopted to describe non-equilibrium phase transition in one-dimensional driven-diffusive systems with open boundaries. To the best of our knowledge the walk models, counterpart of the one-dimensional driven-diffusive systems with periodic boundary conditions, have not been widely studied.\\
In this paper we introduce a two-dimensional walk model and calculate its partition function using a transfer matrix method. We then investigate the critical behaviors of this walk model. Finally, we show that the partition function of the walk model is exactly equal to that of an exactly solvable one-dimensional driven-diffusive system with periodic boundary condition. It is known that this driven-diffusive system can be mapped to a zero-rage process in which a phase transition into a Bose condensate occurs.
\section{The walk model}
\begin{figure}
\includegraphics[width=2.5in]{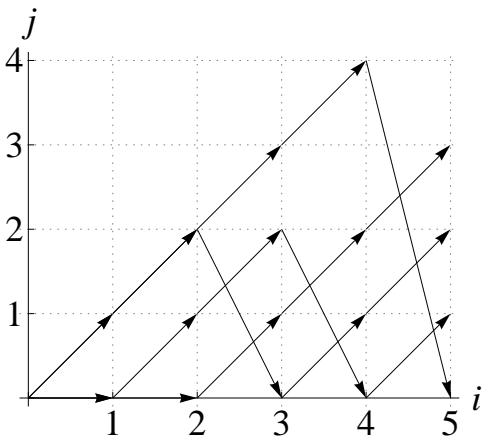}
\caption{\label{Graphics} Four different paths of length $5$ which end up to different heights $0$, $1$, $2$ and $3$.}
\end{figure}

Consider a walk model in which a random walker starts from the origin $(0,0)$ and takes a finite number of steps on $\mathbb{Z}^2_{+}=\{(i,j):i , j\ge 0 \mbox{\; are integers} \}$ according to two forthcoming rules. We assign a weight to each step taken by the random walker. All the paths made by the random walker are weighted. The weight of a given path will be equal to the product of the weights of the consecutive steps in that path. The random walker moves according to the following rules:
\begin{itemize}
\item{ For $i \geq j$ from the lattice site $(i,j)$ to $(i+1,j+1)$ with a weight $\frac{1}{p}$ where $i,j=0,1,2,\cdots,\infty$. This is to be referred to as an upward step.}
\item{ For $i \geq j$ from the lattice site $(i,j)$ to $(i+1,0)$ with a weight $p^{j}$ where $i,j=0,1,2,\cdots,\infty$. For $j\neq0$ ($j=0$) this is to be referred to as a downward (horizontal) step.}
\end{itemize}
We will be finally interested in the paths with fixed number of steps (fixed length) which contain a certain number of downward and horizontal steps (equivalently upward steps); therefore, for our later convenience we introduce an ad hoc fugacity $z$ and change the second rule as follows:
\begin{itemize}
\item{ For $i \geq j$ from the lattice site $(i,j)$ to $(i+1,0)$ with a weight $zp^{j}$ where $i,j=0,1,2,\cdots,\infty$}
\end{itemize}
while keeping the first rule unchanged. As can be seen the random walker does not take any steps in the negative $i$-direction. Let us assume that the random walker starts from the origin $(0,0)$ and takes $N-1$ consecutive steps according to the above mentioned rules. The reason that we have chosen the number of steps as $N-1$ will be clear later. After taking these steps the random walker can get to the lattice site $(N-1,j)$ where $j=0,1,\cdots,N-1$ through different paths. In FIG. \ref{Graphics} we have plotted four different paths of length $5$ according to the above mentioned rules. It is easy to see that there is only one path which ends up at the lattice site $(N-1,N-1)$ and has the weight $1/p^{N-1}$. It can be verified that there are $2^{N-j-2}$ different ways to get to the lattice site $(N-1,j)$ for $j=0,1,\cdots,N-2$. The total weight of the paths which end up at the lattice site $(N-1,j)$ for $j=0,1,\cdots,N-2$ is equal to $z(z+1)^{N-j-2}/p^j$.\\
We define the partition function of the walk model as the sum of the unnormalized weights of different paths consisting of $N-1$ steps.
In what follows we show that the partition function of the model can be obtained using a transfer matrix method. We assume that after $i$ steps the state (position) of the random walker is given by the vector $\vert j \rangle$ where $j$ is the height of the random walker measured from the horizontal axis and can be an integer from $0$ to $i$. These vectors have the following properties:
$$
\begin{array}{l}
\vert j \rangle_{k}=\delta_{j,k}\;\;\mbox{for}\;\;j,k=0,1,\cdots,\infty,\\ \\
\langle j \vert j' \rangle=\delta_{ j , j'}\;\;\mbox{for}\;\;j,j'=0,1,\cdots,\infty,\\ \\
\sum_{j=0}^{\infty}\vert j \rangle \langle j \vert = \mathcal{I}
\end{array}
$$
in which $\mathcal{I}$ is an infinite-dimensional identity matrix. Now we define a transfer matrix $T$ with the following property at each step:
\begin{equation}
\label{TM}
\begin{array}{c}
T\vert j \rangle =zp^{j}\vert 0 \rangle + \frac{1}{p}\vert j+1 \rangle.
\end{array}
\end{equation}
One can introduce the following matrix representation for the transfer matrix $T$:
\begin{equation}
\label{MRT}
T=\left(
\begin{array}{ccccc}
z & zp & zp^2 & zp^3 &\cdots \\
\frac{1}{p} & 0 & 0 & 0 &\cdots\\
0  & \frac{1}{p} &  0  & 0 & \cdots\\
0  & 0 & \frac{1}{p}  & 0 & \cdots\\
\vdots & \vdots & \vdots   & \vdots & \ddots
\end{array} \right).
\end{equation}
In fact the transfer matrix $T$ updates the state (position) of the random walker. The random walker starts from the origin $(0,0)$ which is represented by $\vert 0 \rangle$ at the zero-th step. After taking $N-1$ steps the random walker is in the state $\vert j \rangle$ where $j=0,1,\cdots,N-1$. As we mentioned above, there are different ways (along different paths) to get to the lattice site $(N-1,j)$. The total weight associated with the lattice site $(N-1,j)$ is equal to the sum of the weights associated with the paths that end at that lattice site. It can be verified that:
\begin{equation}
\label{POWERTM}
T^{N-1}\vert 0 \rangle=\sum_{j=0}^{N-2}\frac{z(z+1)^{N-j-2}}{p^j} \vert j \rangle+\frac{1}{p^{N-1}}\vert N-1 \rangle.
\end{equation}
Finally the partition function of the model, which is equal to the sum of the weights associated with the lattice sites $(N-1,j)$ with $j=0,1,\cdots,N-1$, can be easily calculated by multiplying $\sum_{j=0}^{\infty} \langle j \vert$ by (\ref{POWERTM}) from the left:
\begin{eqnarray}
\label{PF1}
Z_{N}(p,z) &=&\sum_{j=0}^{\infty} \langle j \vert T^{N-1} \vert 0 \rangle \nonumber\\
           &=&\sum_{j=0}^{N-2}\frac{z(z+1)^{N-j-2}}{p^j}+\frac{1}{p^{N-1}} \\
           &=&\sum_{i=1}^{N-1} \sum_{j=0}^{N-i-1}
           \left( \begin{array}{c}
           N-j-2\\
           i-1
           \end{array} \right)p^{-j} z^{i}+\frac{1}{p^{N-1}} \nonumber
\end{eqnarray}
in which:
$$
\left( \begin{array}{c} a\\ b \end{array} \right)=\frac{a!}{b!(a-b)!}
$$
is the usual binomial coefficient. As we mentioned above, we are interested in the partition function of the original walk model in a special case that after taking $N-1$ consecutive steps, the random walker has taken exactly $N-M$ upward steps in which $1 \le M \le N$.\\
In order to find the partition function of the model in this case, let us have a closer look at the role of the fugacity $z$. The weight associated with a horizontal or downward movement is proportional to $z$; therefore, the coefficient of $z^{M-1}$ in (\ref{PF1}) is equal to the partition function of the walk model $Z_{N,M}(p)$ which consists of exactly $N-M$ upward steps. One can easily use (\ref{PF1}) to obtain the coefficient of $z^{M-1}$ in $Z_{N}(p,z)$. For $M \neq 1$ the result is:
\begin{equation}
\label{PF2}
Z_{N,M}(p)=\sum_{j=0}^{N-M}
\left( \begin{array}{c}
N-j-2\\
M-2
\end{array} \right) p^{-j}
\end{equation}
and obviously for $M=1$ one finds:
\begin{equation}
Z_{N,1}(p)=\frac{1}{p^{N-1}}.
\end{equation}
The partition function (\ref{PF2}) has a simple explanation in terms of the weighted paths in our walk model. It is the sum of the weights of the paths of length $N-1$ which precisely contain $N-M$ upward steps (or equivalently $M-1$ horizontal and downward steps).\\
As a relevant quantity, one can investigate the mean height of the random walker $\bar{h}$. The probability of being at the height $j$ under the above mentioned conditions is given by:
\begin{equation}
P_{N,M}(j)=
\frac{1}{Z_{N,M}(p)}\left( \begin{array}{c}
N-j-2\\
M-2
\end{array} \right) p^{-j}.
\end{equation}
Now the height of the random walker averaged over all the steps of each walk and over all the walks is given by:
\begin{equation}
\label{MeanH1}
\bar{h}=\sum_{j=0}^{N-M}jP_{N,M}(j)=
-p\frac{\partial \ln Z_{N,M}(p)}{\partial p}.
\end{equation}
It turns out that in thermodynamic limit $N \rightarrow \infty$ with $M=N\rho$, the mean height of the random walker is given by:
\begin{equation}
\label{MeanH2}
\bar{h}\simeq
\left\{
\begin{array}{ll}
N(1-\frac{\rho}{1-p}) \;\;\;\; \mbox{for} \;\; p<1-\rho,\\ \\
\frac{1-\rho}{p-1+\rho} \;\;\;\; \mbox{for} \;\; p>1-\rho.
\end{array}
\right.
\end{equation}
As can be seen, there is a phase transition in the thermodynamic limit from a phase in which the mean height of the random walker is of order $N$ to another phase where it is a constant.\\
In what follows we show that the partition function of the walk model (\ref{PF2}) is exactly equal to that of a driven-diffusive system with periodic boundary conditions. It is known that this driven-diffusive system is equivalent to a zero-range process in which a real-space Bose-Einstein condensation occurs.
\section{An equivalent driven-diffusive system}
In \cite{E96} the author introduces a one-dimensional driven-diffusive system of classical particles with hardcore interactions. The system consists of a particle of type $A$ and $M-1$ particles of type $B$ moving on a one-dimensional lattice of length $N$ with periodic boundary conditions. The particle of type $A$ hops from the lattice site $i$ to $i+1$ with rate $p$ provided the target lattice site is empty. The particles of type $B$ hop from lattice site $i$ to lattice site $i+1$ with rate $1$ provided that the target lattice site is empty. If an empty lattice site is represented by $\emptyset$, then the dynamical rules are simply as follows:
\begin{eqnarray}
\label{ReactionRules}
A \; \emptyset \; &\rightarrow& \; \emptyset \; A \; \mbox{with rate}\; p \nonumber \\
B \; \emptyset \; &\rightarrow& \; \emptyset \; B \; \mbox{with rate}\; 1. \nonumber
\end{eqnarray}
A similar model has also been introduced in \cite{DEHP} and studied in details in \cite{M96}. In this model the particle of type $A$ is called an impurity while the particles of type $B$ are called the normal particles. The normal particles can overtake the impurity. The model can be exactly solved using a matrix product method. It turns out that the phase diagram of the model consists of four different phases. In one of the phases the model presents a shock, i.e. a sharp discontinuity between a region of high density of normal particles and a region of low density. \\
It is shown that the probability distribution of the above mentioned system (the one without overtaking) can also be obtained using a matrix product method \cite{E96}. Let us label the particle of type $A$ with $1$ and the particles of type $B$ with $2,3,\cdots,M-1$ respectively. According to the matrix product formalism, in the steady-state the probability of finding the system in the configuration $\{ {\bf n} \}=\{ n_{1},n_{2},\cdots,n_{M} \}$ in which $n_i$ empty lattice sites lie in front of the $i$th particle is given by:
\begin{equation}
\label{MPPDF}
P(\{ {\bf n} \})=\frac{1}{Z_{N,M}(p)} Tr(D' E^{n_{1}} D E^{n_{2}} \cdots D E^{n_{M}})
\end{equation}
in which the operators $D'$, $D$ and $E$ are associated with the presence of a particle of type $A$, a particle of type $B$ and an empty lattice site respectively. The denominator in (\ref{MPPDF}) is the normalization factor (which is called the canonical partition function) and should be calculated by considering the conservation of number of empty lattice sites i.e. $\sum_{i=1}^{M}n_{i}=N-M$. The expression (\ref{MPPDF}) describes the steady-state of the system provided that operators $D'$, $D$ and $E$ satisfy the following quadratic algebra \cite{E96}:
\begin{eqnarray}
\label{Algebra}
pD'E=D' \\
DE=D. \nonumber
\end{eqnarray}
The partition function $Z_{N,M}(p)$ in (\ref{MPPDF}) has been calculated in \cite{E96} and it turns out that it is exactly equal to that of our walk model given in (\ref{PF2}). In the same reference the critical behaviors of this driven-diffusive system have been investigated. In the thermodynamic limit $L \rightarrow \infty$ and by defining the density of particles $\rho$ as $M=N\rho$ the system undergoes a phase transition from a low-density phase to a high-density phase.
In the low-density phase $p < 1-\rho$ the number of the empty lattice sites in front of the particle of type $A$ is of order $N$. In \cite{E00} the author has shown that the above mentioned driven-diffusive system can be mapped to a zero-range process by converting the particles into boxes and the empty lattice sites into particles. The low-density phase in the driven-diffusive system now corresponds to a Bose-Einstein condensate phase in the equivalent zero-rage process. In contrast, in the high-density phase $p > 1-\rho$ the empty lattice sites are uniformly distributed between the particles \cite{E96}. It is now easy to compare these results with those presented in (\ref{MeanH2}) for the walk model. It turns out that the mean height of the random walker $\bar{h}$ is equivalent to the number of empty lattice sites in front of the particle of $A$ in the driven-diffusive system.\\
Now we explain (from a mathematical point of view) why the partition function of the walk model obtained using the transfer matrix method is equal to that of the driven-diffusive system obtained using the matrix product method. Let us start with the driven-diffusive system and the matrix representation of its quadratic algebra. It can be easily verified that the following infinite-dimensional matrix representation satisfy the algebra (\ref{Algebra}):
\begin{eqnarray}
\label{MatrixRep}
&&D'=\sum_{i=0}^{\infty} \vert 0 \rangle \langle i \vert, \nonumber\\
&&D=\sum_{i=0}^{\infty} p^{i} \vert 0 \rangle \langle i \vert,\\
&&E=\frac{1}{p}\sum_{i=0}^{\infty} \vert i+1 \rangle \langle i \vert \nonumber
\end{eqnarray}
in which $\vert i \rangle_{j}=\delta_{i,j}$ for $i,j=0,1,\cdots,\infty$. Another matrix representation for the algebra (\ref{Algebra}) is given in \cite{BE07}. Although the number of particles in the driven-diffusive system does not change by the dynamical rules, it is much easier to calculate the grand canonical partition function of the driven-diffusive system by defining $z$ as the fugacity of the particles of type $B$. Later we will fix this fugacity by the density of the particles of type $B$. For a system consisting of a single particle of type $A$, the unnormalized weight associated with an arbitrary configuration  is proportional to $Tr(D'X_{\tau_{1}} \cdots X_{\tau_{L-1}})$ in which we have defined $X_{\tau_{i}=0}=E$ and $X_{\tau_{i}=1}=D$ associated with the presence of an empty lattice site and a particle of type $B$ at the lattice site $i$ respectively. Note that in the canonical ensemble one has $\sum_{i=1}^{L-1}\tau_{i}=M-1$.
Now the grand canonical partition function is given by:
\begin{eqnarray}
\label{GCPF1}
Z_{N}(p,z)&=&\sum_{ \{ \tau_{i} \}}z^{M-1}Tr(D'X_{\tau_{1}} \cdots X_{\tau_{L-1}})\nonumber \\
          &=&\sum_{ \{ \tau_{i} \}}       Tr(D'z^{\tau_{1}}X_{\tau_{1}} \cdots z^{\tau_{L-1}}X_{\tau_{L-1}})\\
                    &=&Tr(D'C^{N-1}) \nonumber
\end{eqnarray}
in which we have defined $C=E+zD$. The density of the particles of type $B$ is related to their fugacity through the following relation:
\begin{equation}
\rho=\lim_{N\rightarrow\infty}\frac{z}{N}\frac{\partial \ln Z_{N}(p,z)}{\partial z}.
\end{equation}
Using the matrix representation (\ref{MatrixRep}), the grand canonical partition function (\ref{GCPF1}) can be written as follows:
\begin{eqnarray}
\label{GCPF2}
Z_{N}(p,z)&=&Tr(\sum_{i=0}^{\infty}\vert 0 \rangle \langle i \vert C^{N-1}) \nonumber \\
          &=&\sum_{i=0}^{\infty}\langle i \vert C^{N-1} \vert 0 \rangle.
\end{eqnarray}
Using (\ref{MatrixRep}) it can be seen that the matrix $C$ is exactly equal with the transfer matrix $T$ given in (\ref{MRT}); therefore, the grand canonical partition function of the driven-diffusive system (\ref{GCPF1}) is equal to the partition function of the walk model (\ref{PF1}) where we define the ad hoc fugacity $z$ for the horizontal and downward steps. If we try to impose particle number conservation and adopt a canonical ensemble, then we have to select the coefficient of $z^{M-1}$ in (\ref{GCPF2}) which is exactly (\ref{PF2}) i.e. the partition function of the walk model in the case where the number of upward steps is equal to $N-M$.
\section{Summary and outlook}
It seems that the connection between the two-dimensional walk models and one-dimensional driven-diffusive systems is of a more general validity than a couple of examples studied thus far. The ASEP with open boundaries is an example which has been studied in detail \cite{DEHP}. In \cite{JZ02} the authors have studied the ASEP in the case that its steady-states can be written in terms of a superposition of multiple shocks with random walk dynamics. It turns out that the partition function of the system in this case, is related to that of a walk model consisting of multiple Dyck paths.\\
In this paper we introduced a two-dimensional walk model and calculated its partition function using the transfer matrix method. The paths start from the origin, but in contrast with the perviously introduced walk models, they do not necessarily end on the horizontal axis. We showed that this walk model is closely related to a one-dimensional driven-diffusive system defined on a closed lattice which can be mapped to a zero-range process \cite{E96,E00}. As it is pointed in \cite{E00} a real-space Bose-Einstein condensation can occur in this zero-range process by fine-tuning of the microscopic hopping rates. The zero-range processes have been extensively studied in related literatures \cite{EH05}; therefore, how the zero-range processes and the walk models are related still remains an open question.
\begin{acknowledgments}
The author would like to thank Richard W. Sorfleet for critical reading of this manuscript and helpful suggestions.
\end{acknowledgments}


\begin{thebibliography}{1}
\bibitem{BGR04}  R. Brak, J. de Gier and V. Rittenberg, {\it J. Phys. A: Math. Gen.} {\bf 37} 4303 (2004).

\bibitem{BE04} R. Brak and J. W. Essam, {\it J. Phys. A: Math. Gen.} {\bf 37} 4183 (2004).

\bibitem{BCEP06} R. Brak, S. Corteel, J. Essam, R. Parviainen, A. Rechnitzer, {\it Electron. J. Combin.}
{\bf 13} R108 (2006).

\bibitem{BJJK09} R. A. Blythe, W. Janke, D. A. Johnston and R. Kenna, {\it J. Phys. A: Math. Theor.}  {\bf 42}
325002 (2009).
\bibitem{BE07} R. A. Blythe and M. R. Evans, {\it J. Phys. A: Math. Theor.} {\bf 40} R333 (2007).
\bibitem{JZ01} F. H. Jafarpour and S. Zeraati, {\it Phys. Rev. E} {\bf 81} 011119 (2010).
\bibitem{JZ02} F. H. Jafarpour and S. Zeraati, {\it Phys. Rev. E} {\bf 82} 041133 (2010).
\bibitem{BE03} R. A. Blythe and M. R. Evans, {\it Braz. J. Phys.} {\bf 33} 3 (2003).
\bibitem{DEHP} B. Derrida, M.R. Evans, V. Hakim, V. Pasquier, {\it J. Phys. A} {\bf 26} 1493 (1993).
\bibitem{M96} K. Mallick, {\it J. Phys. A: Math. Gen.} {\bf 29} 5375 (1996).
\bibitem{E96} M. R. Evans, {\it Europhys. Lett.} {\bf 36} 13 (1996).
\bibitem{E00} M. R. Evans, {\it Braz. J. Phys.} {\bf 30} 1 (2000).
\bibitem{BOR08} R. Brak, A. L. Owczarek and A. Rechnitzer {\it J. Math. Chem.} {\bf 45} 39 (2008).
\bibitem{EH05} M. R. Evans and T. Hanney, {\it J. Phys. A: Math. Gen.} {\bf 38} R195 (2005)
\end{thebibliography}
\end{document}